\documentclass[10pt,twocolumn,prb,aps,floatfix,superscriptaddress,longbibliography]{revtex4-1}

\usepackage{color}
\usepackage{graphicx}
\usepackage{amsthm}
\usepackage{amsmath}
\usepackage{amssymb}
\usepackage{enumerate}
\usepackage{placeins}

\newcommand{\subfigimg}[3][,]{%
  \setbox1=\hbox{\includegraphics[#1]{#3}}% Store image in box
  \leavevmode\rlap{\usebox1}% Print image
  \rlap{\hspace*{1pt}\raisebox{\dimexpr\ht1+5pt}{#2}}% Print label
  \phantom{\usebox1}% Insert appropriate spcing
}

\DeclareMathOperator{\Tr}{Tr}

\newcommand{\ket}[1]{\vert #1 \rangle}
\newcommand{\eket}[1]{\bigl \vert #1 \bigr \rangle}
\newcommand{\R}{\boldsymbol{R}}
\newcommand{\Rt}{\tilde{\R}}

\newcommand{\li}{\ell_{\rm int}}

\newcommand{\ebra}[1]{\bigl \langle #1 \bigr \vert}
\newcommand{\eexp}[1]{\bigl \langle #1 \bigr \rangle}
\newcommand{\figref}[1]{Fig.~\ref{#1}}
\renewcommand{\vec}[1]{\boldsymbol{#1}}
\newcommand{\ren}{R\'{e}nyi~}
\newcommand{\Eqref}[1]{Eq.~\eqref{#1}}

\begin{document}

\title{Spatial entanglement entropy in the ground state of the Lieb-Liniger model}

\author{C. M. Herdman}
\email{cherdman@uwaterloo.ca}
\affiliation{Institute for Quantum Computing, University of Waterloo, Ontario, N2L 3G1, Canada}
\affiliation{Department of Physics \& Astronomy, University of Waterloo, Ontario, N2L 3G1, Canada}
\affiliation{Department of Chemistry,  University of Waterloo, Ontario, N2L 3G1, Canada}

\author{P.-N. Roy}
\affiliation{Department of Chemistry, University of Waterloo, Ontario, N2L 3G1, Canada}

\author{R. G. Melko}
\affiliation{Department of Physics \& Astronomy, University of Waterloo, Ontario, N2L 3G1, Canada}
\affiliation{Perimeter Institute for Theoretical Physics, Waterloo, Ontario N2L 2Y5, Canada}

\author{A. Del Maestro}
\affiliation{Department of Physics, University of Vermont, Burlington, VT 05405, USA}

\begin{abstract}
We consider the entanglement between two spatial subregions in the Lieb-Liniger model of bosons in one spatial dimension interacting via a contact interaction. 
Using ground state path integral quantum Monte Carlo we numerically compute the \ren entropy of the reduced density matrix of the subsystem as a measure of entanglement. Our numerical algorithm is based on a replica method previously introduced by the authors,
which we extend to efficiently study the entanglement of spatial subsystems of itinerant bosons. 
We confirm a logarithmic scaling of the \ren entropy with subsystem size that is expected from conformal field theory, and compute the non-universal subleading constant for interaction strengths ranging over two orders of magnitude. In the strongly interacting limit, we find agreement with the known free fermion result.
\end{abstract}

\maketitle

% ---------------------------------------------------------------------------------
% Introduction
\section{Introduction}
% ---------------------------------------------------------------------------------

The Lieb-Liniger model of $\delta$-function interacting bosons in the one dimensional (1D) spatial continuum~\cite{Lieb1963a,Lieb1963} is one of only a handful of quantum-many body systems with pairwise interactions where the ground state wavefunction is known exactly.  In addition to its theoretical importance and connection to the Tonks-Girardeau gas~\cite{Tonks:1936eq, Girardeau1960a} that exhibits Bose-Fermi correspondence, the Lieb-Liniger model can be experimentally probed in quasi-one dimensional systems of ultracold atoms \cite{Kinoshita2005,Haller:2009jr, Meinert:2015dx} and used to model clusters of bosonic solvent particles, doped with a molecular rotor~\cite{Wairegi2014,Farrelly2016}. These experimental realizations of Lieb-Liniger systems have lead to a renewed interest in its physical properties, with a flurry of recent works developing a high-precision understanding of its correlations (both in real and momentum space) and excitation spectrum \cite{Astrakharchik:2015ww,Boeris:2015un,Xu:2015go, Choi:2015fn, Bertaina:2016vx}. However, the degree to which those correlations are non-classical, as reflected in the entanglement structure of the ground state, has not been fully characterized. 

Entanglement is a fundamental property of all quantum systems that is known to be a resource for quantum information processing \cite{Horodecki2009,Amico2008}. The structure and finite size scaling of entanglement can reveal features of quantum phases of matter and phase transitions,~\cite{Osterloh2002,Vidal2003,Levin2006a,Kitaev2006b,Hastings2007} and has implications for the simulation of quantum systems on classical computers \cite{Verstraete2006a,Schuch2008a}.  While its naive measurement in a $N$-body system would seem to require access to the exponentially large density matrix corresponding to its quantum state, field theoretic\cite{Calabrese2004}, algorithmic~\cite{Hastings2010}, and experimental \cite{Daley:2012bd, Islam:2015cm} advances have led to the ability to compute and measure it using the expectation value of local operators.

This has led to a number of studies focusing on entanglement in lattice models with insulating degrees of freedom \cite{Melko2010,Humeniuk2012b,Broecker2014,Wang2014}.  In contrast, much less is known about 
the entanglement properties of quantum fluids \cite{Calabrese:2009co}. Such continuum systems pose significant theoretical challenges due to their formally infinite Hilbert spaces\cite{Adesso:2007eq} and the indistinguishability and itinerance of their constituent particles \cite{Dowling:2006kq}.  For non-interacting gases, studies of the bipartite spatial entanglement \cite{Calabrese2011a,Calabrese2011} have confirmed the logarithmic finite size scaling predicted by conformal field theory. For interacting particles in the continuum, progress has been made using Monte Carlo methods, including variational studies of fermions~\cite{McMinis2013,Tubman2012} the entanglement of bosons under a particle partition~\cite{Herdman2014a,Herdman:2015gx} and the spatial entanglement of small systems of $N=4$ bosons~\cite{Herdman2014}. Additionally, continuous matrix product states methods have been used to study the entanglement of the infinite half chain of the Lieb-Liniger model as a function of bond dimension~\cite{Rincon2015}.

In this paper we introduce a quantum Monte Carlo technique which employs the ``ratio method'' \cite{Hastings2010,Humeniuk2012b} enabling the unbiased calculation of spatial partition entanglement in the ground state of the Lieb-Liniger model with large $N$.  This algorithm is generally applicable to systems of itinerant non-relativistic bosons in any spatial dimension $D$ with the canonical Hamiltonian:
\begin{equation}
    H = \sum_{i=1}^N \left({ - \frac{\hbar^2}{2m_i} \nabla_i^2 + U_i}\right) + \sum_{i<j} V_{ij},
\label{eq:Ham}
\end{equation}
where $U_i$ is an external and $V_{ij}$ an interaction potential.  By performing large scale simulations of the Lieb-Liniger model with $N$ up to 32, we are able to confirm 
the leading order logarithmic finite size scaling of the entanglement entropy, recovering the expected value of $c=1$ for the central charge of the underlying conformal field theory. We 
observe $N$-independent non-universal scaling corrections which decrease monotonically with increasing interactions, yielding the expected free Fermion result in the strongly interacting limit. 

The rest of this paper is organized as follows.  We begin by introducing the \ren entanglement entropy and discuss what is currently known about its finite size scaling in critical 1D systems. After describing the relevant details of the Lieb-Liniger model under consideration where $U_i = 0$ and $V_{ij} \propto \delta(x_i-x_j)$ in Eq.~\eqref{eq:Ham} we introduce and benchmark a quantum Monte Carlo method able to measure the entanglement entropy.  We present our scaling results and finally discuss the potential of using this method to further probe so-called \emph{unusual} corrections to scaling \cite{Cardy2010,Sahoo:2016ch}.

% ---------------------------------------------------------------------------------
% Entanglement Entropy}
 \section{Entanglement entropy in critical one dimensional systems}
 \label{sec:EECFT}
% ---------------------------------------------------------------------------------

\begin{figure}
\begin{center}
\includegraphics[width=0.7\columnwidth]{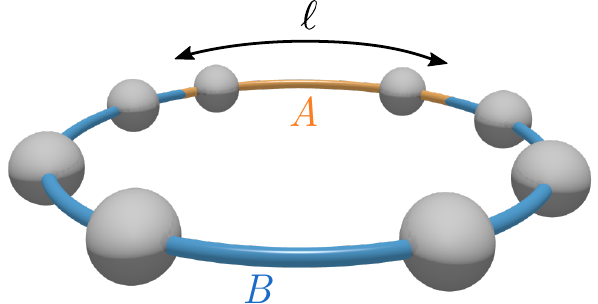}
\end{center}
\caption{(Color online) Spatial bipartition of a system of $N$ itinerant indistinguishable particles in one dimension of length $L$ with periodic boundary conditions into two subregions $A$ and $B$, where $A$ has length $\ell$. Particles can dynamically move between subregions.} 
\label{fig:spatialBipartition}
\end{figure}

We consider the bipartite entanglement between two spatial subregions of the ground state $\ket{\Psi_0}$ of a critical one dimensional system as shown in \figref{fig:spatialBipartition}. A spatial bipartition defines two intervals $A$ and $B$ with the reduced density matrix of the $A$ subsystem, $\rho_A$, defined as
\begin{equation}
\rho_A \equiv \Tr_B \eket{\Psi_0}\ebra{\Psi_0}
\end{equation}
where $\Tr_B$ indicates a partial trace over all degrees of freedom in $B$.
The entanglement between the subsystems may be quantified by the \ren entropy of $\rho_A$:
\begin{equation}
S_{\alpha}\left[ \rho_A \right] \equiv \frac{1}{1-\alpha} \log \bigl({ {\rm Tr} \rho_A^{\alpha} }\bigr), \label{eq:renyi}
\end{equation}
where $\alpha$ is the \ren index. For $\alpha \rightarrow 1$ the \ren entropy is equivalent to the von Neumann entropy: $-\mathrm{Tr}\, \rho_A \log \rho_A$. 

The entanglement entropy (EE) is bounded from above by the logarithm of the dimension of the Hilbert space of the subsystem. For itinerant particles in the spatial continuum, any non-trivial partition always has an infinite dimensional Hilbert space, and therefore no upper bound on the entanglement entropy would seem to exist. However, for a system with a local Hamiltonian, finite-energy states are expected to have finite entanglement between $A$ and $B$~\cite{Eisert2002,Eisert2003}.

Moreover, the ``area law'' of entanglement entropy states that the bipartite entanglement of a gapped 1D system should be a non-universal constant, independent of the subsystem size~\cite{Srednicki1993,Hastings2007a,Hastings2007,Eisert2010}. In contrast, critical quantum systems in 1D described by conformal field theory  (CFT) are know to have an entanglement entropy the diverges \emph{logarithmically} with subsystem size $\ell$ in the thermodynamic limit~\cite{Holzhey1994,Vidal2003,Calabrese2004},
\begin{equation}
S^{\rm{CFT}}_\alpha \left(\ell\right) \simeq \frac{c}{6} \left(1 + \frac{1}{\alpha}\right) \log \ell + \dots, \label{eq:SCFT}
\end{equation}
where $c$ is the central charge of the CFT and $\alpha$ is the \ren index.  Therefore, whereas for gapped systems the leading order (constant) scaling of the entanglement entropy is determined by the microscopic physics at the interface, for critical systems the leading order scaling is universal and determined by the effective low energy field theory.

For a critical 1D ground state in a finite sized system of length $L$ with periodic boundary conditions, the interval $\ell$ in \Eqref{eq:SCFT} is replaced by the chord length $D(L,\ell)$:
\begin{equation}
D(L,\ell) \equiv \frac{L}{\pi} \sin  \left(\pi \frac{\ell}{L} \right)
\end{equation}
such that the scaling of $S_\alpha$ due to the CFT is~\cite{Calabrese2004,Cardy2010}
\begin{align}
S^{\rm{CFT}}_\alpha \left(L,\ell\right) = \frac{c}{6} \left(1 + \frac{1}{\alpha}\right) \log &\Bigl[ D\left(L,\ell\right) \Bigr] + c_\alpha \notag \\
&+\mathcal{O} \left( \ell^{-p_{\alpha}}\right), 
\label{eq:FSCFT}
\end{align}
where $c_\alpha$ is a non-universal constant and $p_\alpha$ is the exponent of the leading order corrections. The power-law corrections to this scaling can include non-universal terms due to irrelevant operators in the bulk of the subsystem as well as universal terms due to relevant operators~\cite{Laflorencie2006,Calabrese2010c,Cardy2010}.  Previous numerical studies of these corrections to scaling have been undertaken for 1D XXZ lattice spin models \cite{Laflorencie2006, Calabrese2010b, Calabrese2010c,Calabrese2011a} as well as other discrete symmetry systems including the Ising, Blume-Capel, and the three-state Potts models \cite{Xavier:2012en}, dipolar bosons on a lattice \cite{Dalmonte:2011fm} and Fermi liquids \cite{Swingle:2013gh}.  A common feature of these studies is the observation of spatial $2k_{\mathrm{F}}$-like oscillations in the subleading corrections. Their origins, along with uncertainties on the model, symmetry and interaction dependence of the \ren index dependent power $p_\alpha$ in Eq.~\eqref{eq:FSCFT}, are not fully understood.  Thus, performing a careful scaling analysis of the entanglement entropy in the Lieb-Liniger model where ultraviolet effects (due to a lattice) are not present may provide new insights into these issues.  Moreover, apart from being purely of theoretical interest, a detailed understanding of EE scaling corrections may be essential to distinguish different theories with the same central charge without having to resort to studying disjoint intervals \cite{Alba:2010bk}.
%
% ------------------------------------------------------------------------------- 
% ------------------------------------------------------------------------------- 
%

% ---------------------------------------------------------------------------------
% Luttinger-Liquids}
\section{Lieb-Liniger Model}
% ---------------------------------------------------------------------------------

The Lieb-Liniger model describes $N$ spinless non-relativistic bosons interacting with a contact interaction in one dimensional continuous space~\cite{Lieb1963a,Lieb1963} with Hamiltonian:
\begin{equation}
    H = -\lambda \sum_{i=1}^N \frac{d^2}{d x_i^2} + g\sum_{i<j} \delta\left(x_i-x_j\right), 
    \label{eq:HLieb}
\end{equation}
where $\lambda \equiv \hbar^2/2m$ and $g$ is the interaction strength with dimensions of energy $\times$ length. We consider only repulsive interactions $g\geq0$ and as $g\rightarrow + \infty$ the Tonks-Girardeau~\cite{Tonks:1936eq,Girardeau1960a} gas of impenetrable bosons is recovered. Here we consider a finite system of length $L$ with periodic boundary conditions, and define the number density $n\equiv N/L$. There are two relevant short distance length scales: the interparticle separation $\ell_0 \equiv 1/n$ and the interaction length scale $\ell_{\rm{int}} \equiv 2\lambda/g$.  It is useful to parameterize finite interactions using a single dimensionless parameter $\gamma \equiv \ell_0/\ell_{\rm{int}}$, which, as mentioned in the introduction can be experimentally tuned in ultracold Bose gases confined in quasi-1D optical traps.\cite{Kinoshita2005}  

The low energy physics of the Lieb-Liniger model is described by Tomonaga-Luttinger liquid (TLL) theory~\cite{Tomonaga1951,Luttinger1963,Mattis1965,Haldane1981}. Tomonaga-Luttinger liquids are critical quantum phases whose non-universal properties are characterized by a single energy scale $v$ and a single dimensionless Luttinger parameter $K$ which determines the power-law decay of correlation functions. Due to the conformal invariance of TLL theory, the low energy physics of the Lieb-Liniger model is described by a CFT with universal central charge $c=1$~\cite{Cazalilla2004,Cazalilla2011}.  Consequently, the correlation functions of the ground state of the Lieb-Liniger model decay as non-universal power-laws whose exponents depend on $\gamma$ since the effective Luttinger parameter $K(\gamma)$ is a non-trivial function of $\gamma$. On the other hand, since a TLL is described by a $c=1$ CFT, the leading order scaling of the spatial entanglement entropy is expected to be of the universal CFT form  given in \Eqref{eq:FSCFT} with $c=1$. The non-universal constant $c_\alpha$ and power-law corrections are expected to depend $\gamma$.  
For the rest of this paper we will consider only $\alpha=2$.
Then, the CFT asymptotic scaling of the 2nd \ren entanglement entropy for the ground state of the Lieb-Liniger model at fixed interaction strength $\gamma$ may be written as
\begin{align}
    S^{\rm{LL}}_2 \left(N,\ell\right) &= \frac{1}{4} \log \Bigl[ 2 \pi n D\left(N,\ell\right) \Bigr]+ c_2+ \mathcal{O} \left( \ell^{-p_{2}}\right),\label{eq:FSCFTc1} 
\end{align}
where we now write the chord length as a function of $N$ and $\ell$ at fixed density:
\begin{equation}
D(N,\ell) \equiv \frac{N}{\pi n} \sin  \left(\pi n\frac{\ell}{N} \right).
\end{equation}
We have chosen this definition of the subleading constant $c_2$ to be consistent with existing literature where it was calculated for the ground state of free fermions in the 1D spatial continuum, and shown to be equal to the subleading constant of the $XY$ lattice spin model~\cite{Calabrese2011a,Calabrese2011}.  As it is known that the Lieb-Liniger (LL) model maps onto free fermions in the strongly interacting Tonks-Girardeau limit $\gamma \rightarrow \infty$ we expect: 
\begin{equation}
c_2^{\rm{LL}}\left(\gamma = \infty\right) = c_2^{\rm{FF}} \simeq 0.404049.
\end{equation}

% ---------------------------------------------------------------------------------
% Quantum Monte Carlo Method}
 \section{Quantum Monte Carlo}
 \label{sec:QMC}
% ---------------------------------------------------------------------------------

 \subsection{Path integral ground state Monte Carlo}
% ---------------------------------------------------------------------------------

 To compute the EE of the ground state of the Lieb-Liniger model under a spatial bipartition, we use a path integral ground state quantum Monte Carlo (PIGS) method \cite{Ceperley1995,Sarsa2000} which provides unbiased access to ground state expectation values through imaginary time projection: 
\begin{equation}
\eexp{ \hat{\mathcal{O}} } = \lim_{\beta \rightarrow \infty}\frac{ \ebra{\Psi_{\mathrm{T}}}e^{-\beta H/2} \hat{\mathcal{O}} e^{-\beta H/2} \eket{\Psi_{\mathrm{T}}}}{\ebra{\Psi_{\mathrm{T}}}e^{- \beta H} \eket{\Psi_{\mathrm{T}}}} \label{eq:O_PIGS}
\end{equation}
where $ \hat{\mathcal{O}}$ an observable and $\ket{\Psi_{\mathrm{T}}}$ is a trial wave function (we now choose units with $\hbar=1$).  The Monte Carlo sampling of \Eqref{eq:O_PIGS} is done over a configuration space comprising imaginary time worldlines of $N$ bosons in one spatial dimension. Using a discrete imaginary time representation, we approximate the propagator as the product of short time propagators:
\begin{equation}
e^{-\beta H} \simeq \left( \rho_\tau \right)^P
\end{equation}
with  $P \equiv \mathrm{int}[\beta/\tau]$ which is exact in the limit $\beta\to\infty$. 
The imaginary time worldline configurations in the position basis is represented such that each imaginary time slice is described by a state $\ket{\R}$, where 
\begin{equation}
\R = \{r_0,\ldots,r_{N-1}\}
\end{equation}
is a vector of length $N$ describing the position of all particles in continuous space (beads) at that time slice. The short time propagator $\rho_\tau$ is approximately decomposed into the product of the free particle propagator, $\rho_0$ which can be sampled exactly, and an interaction propagator, $\rho_{\rm int}$,
\begin{align}
 \rho_\tau (\R,\R') &=  \ebra{\R} e^{-\tau H} \eket{\R'} \notag \\
 &\simeq \rho_0(\R,\R';\tau,\lambda) \rho_{\rm{int}}(\R,\R';\tau)
\end{align}
where $\rho_0(\R,\R';\tau,\lambda)$ is the free $N$ particle propagator:
\begin{equation}
    \rho_0(\R,\R';\tau,\lambda) \equiv \prod_{j=0}^{N-1} \rho_0 \left(r_j-r_j',\tau,\lambda \right)
\end{equation}
with
\begin{equation}
    \rho_0 \left(\Delta x,\tau,\lambda\right) = \frac{e^{-\Delta x^2/4\lambda \tau}}{2 \sqrt{\pi \lambda \tau}} .
\end{equation}
Due to the infinitely short-ranged nature of interactions in the Lieb-Liniger model (\Eqref{eq:HLieb}), the short time propagator must be sampled using 
a pair-product decomposition \cite{Ceperley1995} which employs the exact two-body propagator for $\delta$-function interacting bosons~\cite{Gaveau1986,Casula2008,Manoukian1989}:
\begin{equation}
 \rho_{\rm{int}} \left(\R, \R';\tau\right) \simeq \prod_{j \neq k} W_{\rm int} \left(r_j-r_k,r'_j-r'_k;\tau\right).
\end{equation} 
Here $W_{\rm int}$ is a weight that takes into account the pairwise interactions, and only depends on the relative separation of each pair across a time-slice. The explicit form of $W_{\rm int}$ for the Lieb-Liniger model is given in the Appendix of Ref.~[\onlinecite{Herdman:2015gx}].  Finally, the weight of a segment of the imaginary-time path $\{\R_{m},\R_{m+1},\dots,\R_{m+M}\}$ of length $M\tau$ is
\begin{align}
    W\left(\R_m,\dots,\R_{m+M}\right) &= \rho_\tau (\R_{m},\R_{m+1}) 
    \rho_\tau (\R_{m+1},\R_{m+2})
    \notag \\
&\quad \times
\cdots\rho_\tau (\R_{m+M-1},\R_{m+M}).
\end{align}
Updates to the interior of these segments can be done with conventional path integral Monte Carlo updates~\cite{Ceperley1995}.

%
% ------------------------------------------------------------------------------- 
\begin{figure}
\begin{center}
\includegraphics[width=1.0\columnwidth]{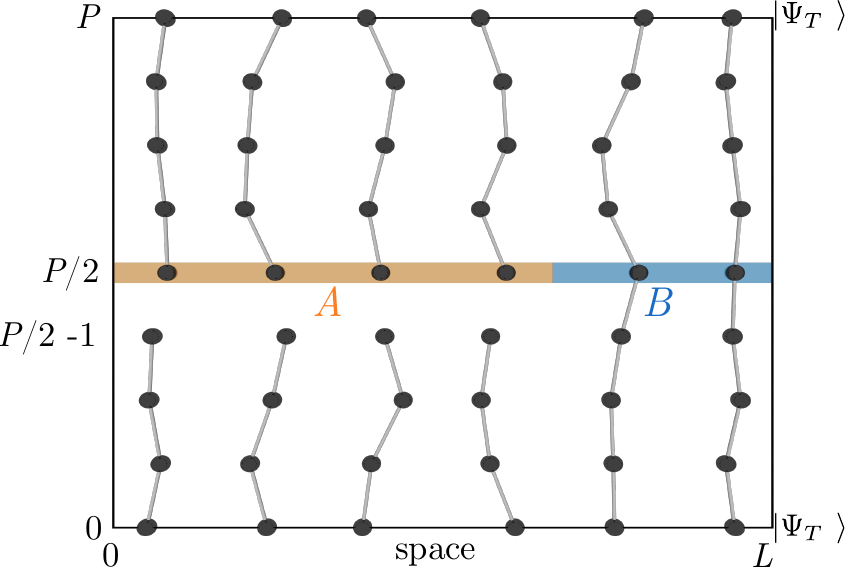}
\end{center}
\caption{(Color online) Broken worldline configuration used to measure the SWAP operator.  $N=6$ worldlines formed by projecting $|\Psi_T\rangle$ are continuous through the $B$ spatial subregion, but are cut between slices $P/2-1$ and $P$ in subregion $A$.}
\label{fig:brokenWorldlines}
\end{figure}
% ------------------------------------------------------------------------------- 
%

 \subsection{The SWAP method}
 \label{sec:SWAP}
% ---------------------------------------------------------------------------------

 Although the \ren EE is not a conventional observable, previous literature has demonstrated that it can be successfully computed via Monte Carlo methods, by writing $\Tr \rho_A^\alpha$ as an expectation value in a replicated configuration space, where multiple identical copies of the same physical system are sampled simultaneously~\cite{Hastings2010}. For example, $S_2$ is accessible via QMC by sampling two identical, non-interacting replicas of the physical system under consideration. We now review this previously introduced replica (or SWAP) method \cite{Herdman2014a,Herdman2014} in the context of the Lieb-Liniger model under consideration here.  

For continuous space path integral Monte Carlo, the replica method can be utilized by sampling an ensemble of imaginary-time worldlines that are \emph{broken} at the center of both paths corresponding to the $A$ subsystem consisting of $n$ particles~\cite{Herdman2014a,Herdman2014} as shown in \figref{fig:brokenWorldlines}.  For a spatial partition, a configuration $\R$ can be dynamically partitioned into sets of particles in the $A$ and $B$ subsystems such that
\begin{equation}
\R = \R_A \cup \R_B 
\end{equation}
where $\R_{A}$ ($\R_{B}$) is vector of positions of particles in the $A$ ($B$) subsystems. The weight for these broken paths is
\begin{align}
\Psi_{\mathrm{T}}\left(\R_0 \right) W\left(\R_0,\dots,\R_{P/2-1}\right)&  \rho_B (\R_{P/2-1},\R_{P/2}) \notag\\
\times W\left(\R_{P/2},\dots,\R_P\right) &\Psi_{\mathrm{T}}\left(\R_P \right)
\end{align}
where $\rho_B$ is the symmetrized reduced propagator for the $B$ subsystem:
\begin{align}
 \rho_B \left(\R,\R'\right) \equiv& \frac{\left(N-n_B\right)!}{N!} \sum_{\R_{n_B}} \sum_{\mathcal{P}\left(\R'_B\right)} \notag \\ 
 \rho_0(\R_{n_B},&\mathcal{P}\left(\R'_B\right);\tau,\lambda)\rho_{\rm{int}}(\R_{n_B},\mathcal{P}\left(\R'_B\right);\tau), 
    \label{eq:BrokenWeights}
\end{align}
$n_B$ is the number of particles in $\R_B'$, $\R_{n_B}$ is one subset of $n_B$ particles of $\R$, the first sum is over all such subets, and the 2nd sum is over all permutations of $\R'_B$. The contribution to the weight from the trial wavefunction is $\Psi_T(\R) \equiv \langle \R | \Psi_T \rangle$.  The key point here is that in \Eqref{eq:BrokenWeights} there is no kinetic propagator connecting the particles in the $A$ subsystem of $\R_{P/2-1}$ to the next time slice $\R_P$.

The estimator for $\Tr \rho_A^2$ is related to the expectation value of the short-imaginary-time propagator which connects the broken worldlines across the replicas. We define the reduced propagator for the $A$ subsystem as
\begin{equation}
\rho_A \left(\R,\R'\right) \equiv \frac{\rho_\tau \left(\R,\R'\right)}{\rho_B \left(\R,\R'\right)}.
\end{equation}
The replica approach then requires sampling two independent, non-interacting copies of the system, each with a weight given by \Eqref{eq:BrokenWeights}, with broken worldlines at the $P/2$ time slice. The estimator within this replicated configuration space is 
\begin{equation}
\Tr \rho_A^2 = \frac{\left\langle \rho_A^{\mathrm{SWAP}} \right \rangle_A}{\left\langle \rho_A^{\mathrm{DIR}} \right \rangle_A}, \label{eq:BrokenEstimator}
\end{equation}
where $\rho_A^{\mathrm{DIR}}$ and $\rho_A^{\mathrm{SWAP}}$ are the reduced propagators for the A subsystems which connect the broken beads to the same and other replica, respectively. The notation $\langle \cdots\rangle_A$ indicates an ensemble average over worldline configurations with open paths in region $A$. Note that the denominator in \Eqref{eq:BrokenEstimator} is a normalization factor that is required due to the configuration space of open paths.

 \subsection{Ratio method}
% ---------------------------------------------------------------------------------

A major obstacle for using the basic SWAP method presented in section \ref{sec:SWAP} is that in general both the numerator and denominator  of \Eqref{eq:BrokenEstimator} decay exponentially with the size of the subsystem $A$. This is a general problem encountered in all SWAP Monte Carlo based approaches. Ultimately the basic SWAP estimator is expected to decay exponentially with the amount of entanglement between the subsystems, and, with the exception of gapped 1D systems, this entanglement is expected to grow (at least logarithmically) with the subsystem size. In the context of continuous-space worldline Monte Carlo, the exponential decay of the components of the estimator arises from the product of Gaussian factors from the free particle propagator.

\begin{figure}[t]
\begin{center}
\includegraphics[width=1.0\columnwidth]{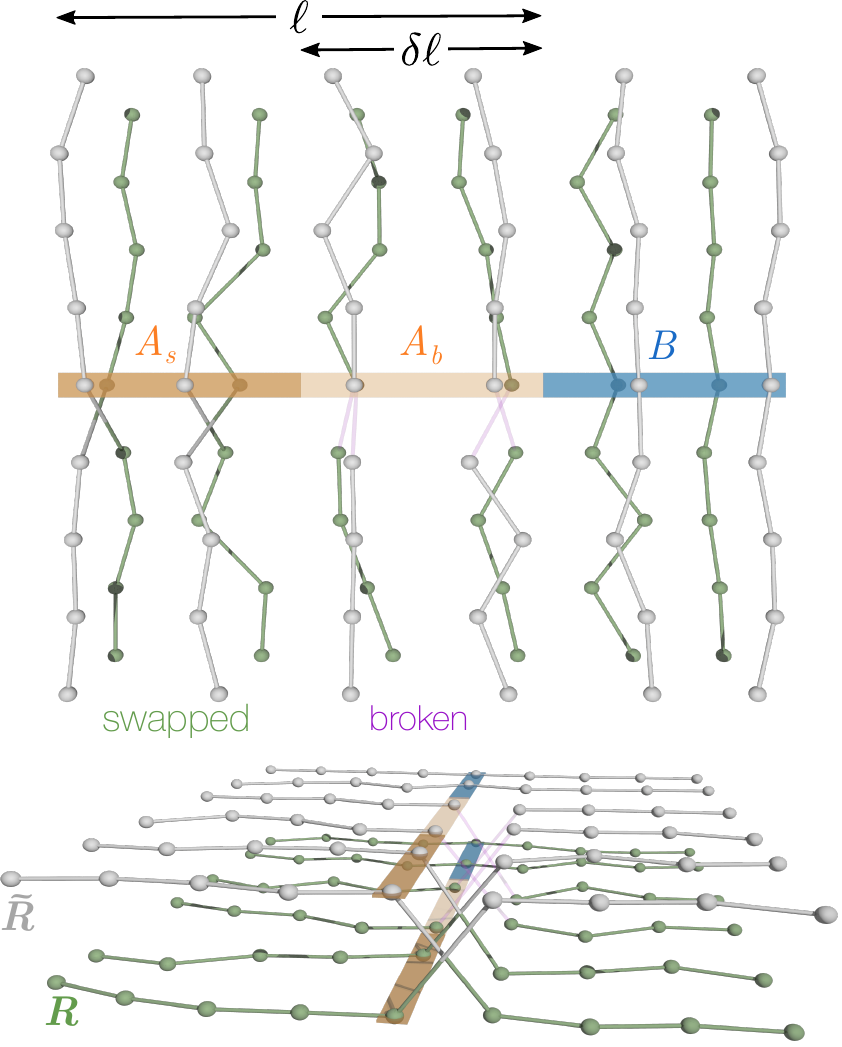}
\end{center}
\caption{(Color online) Top and side view schematic of the replicated configuration space of $N = 6$ bosons showing the decomposition of subregion $A = A_s \cup A_b$ required to efficiently compute the 2nd \ren entropy for large subregion size $\ell$. Worldlines which pass through region $A_s$ at time-slice $P/2$ are connected via an insertion of the short-distance propagator $\rho_\tau$ between replicas (forming part of the ensemble), while those in ration $A_b$ remain broken.  A translucent connection between $\vec{R}$ and $\vec{\tilde{R}}$ is used to indicate which \emph{broken} beads are connected during the SWAP estimation procedure. For the case shown here we have $K=2$ and $\delta \ell = \ell / 2$. }
\label{fig:swapReplicas}
\end{figure}

A successful route for circumventing this problem was developed in the context of lattice models \cite{Hastings2010} where improved performance is obtained by building up the desired estimator from a ratio of estimators for smaller spatial subregions. To see this, we first decompose partition $A$ into contiguous subregions $A_s$ and $A_b$ (s$\to$swapped, b$\to$broken) such that $A = A_s \cup A_b$. Now we define a new configuration space where the two replicas are connected via an imaginary-time propagator in region $A_s$ but the wordlines remain broken in region $A_b$ (see \figref{fig:swapReplicas}). 
%
% ------------------------------------------------------------------------------- 
% ------------------------------------------------------------------------------- 
%
The weights for this ensemble are
\begin{align}
    &\Psi_{\mathrm{T}}\left(\R_0 \right) W\left(\R_0,\dots,\R_{P/2-1}\right) \rho_{A_s}\left(\R_{P/2-1},\Rt_{P/2}\right)\notag \\
&\times  \rho_B (\R_{P/2-1},\R_{P/2}) W\left(\R_{P/2},\dots,\R_P\right) \Psi_{\mathrm{T}}\left(\R_P \right) \notag\\
    & \times \Psi_{\mathrm{T}}\left(\Rt_0 \right) W\left(\Rt_0,\dots,\Rt_{P/2-1}\right)\rho_{A_s}\left(\Rt_{P/2-1},\R_{P/2}\right)\notag \\
    &\times \rho_B (\Rt_{P/2-1},\Rt_{P/2}) W\left(\Rt_{P/2},\dots,\Rt_P\right) \Psi_{\mathrm{T}}\left(\Rt_P \right),
\end{align}
and we indicate statistical averages in this ensemble via $\langle \cdots\rangle_{A_b;A_s}$. The estimator for $\Tr \rho_A^2$  is formed from a product of estimators over two different ensembles:
\begin{equation}
\frac{\left\langle \rho_A^{\mathrm{SWAP}} \right \rangle_A}{\left\langle \rho_A^{\mathrm{DIR}} \right \rangle_A} = \frac{\left\langle \rho_{A_s}^{\mathrm{SWAP}} \right \rangle_{A_s}}{\left\langle \rho_{A_s}^{\mathrm{DIR}} \right \rangle_{A_s}} \frac{\left\langle \rho_{A_b}^{\mathrm{SWAP}} \right \rangle_{A_b;A_s}}{\left\langle \rho_{A_b}^{\mathrm{DIR}} \right \rangle_{A_b;A_s}}.
\end{equation} 
The improved performance of this estimator is due to the reduced size of the ``broken" region for which the imaginary-time propagator is measured in each individual statistical average. However, this gain is achieved at the cost of performing an additional simulation over a different ensemble.

This approach can be generalized in a straightforward manner by partitioning $A$ into $K$ regions such that
\begin{equation}
A = A_1 \cup A_2 \cup \dots \cup A_{K}.
\end{equation}
and using $K$ independent simulations using different ensembles; at the $k^{\rm{th}}$ step, $A_s = A_1 \cup \dots \cup A_{k-1}$ and $A_b = A_{k}$, providing an ensemble to compute the $k^{\rm{th}}$ ratio $\Phi_k$, defined as
\begin{equation}
\Phi_k \equiv \frac{\left\langle \rho_{A_k}^{\mathrm{SWAP}} \right \rangle_{A_k; \bigcup_{k'=1}^{k-1}A_{k'}}}{\left\langle \rho_{A_k}^{\mathrm{DIR}} \right \rangle_{A_k; \bigcup_{k'=1}^{k-1}A_{k'}}}.
\label{eq:Phik}
\end{equation}
Thus the estimator for $\Tr \rho_A^2$ is then a product of estimators from $K$ simulations:
\begin{equation}
\Tr \rho_A^2 = \prod_{k=1}^{K}\Phi_k.
\end{equation}
Another straightforward generalization of this approach is to include updates which change $A_s$ and $A_b$ (i.e. allowing $A_s$ to grow and shrink during a single simulation). Such an approach would allow $\Tr \rho_A^2$ to be computed from a single simulation, taking advantage of the efficiency of the ratio method (e.g see  Ref.~[\onlinecite{Humeniuk2012b}]).

 \subsection{Updates for the ratio method}
% ---------------------------------------------------------------------------------

 To ergodically sample the configuration space used in the ratio method, we use updates that can be grouped into four general categories: closed segment updates, open segment updates, break-connect updates and cross segment updates. Closed segment updates address closed  worldline pieces entirely within a single replica and can be performed within the conventional path integral Monte Carlo (PIMC) scheme~\cite{Ceperley1995}. Open segment updates address imaginary time segments which are open at one end and remain open throughout the update.  These updates are performed in tandem with those used for conventional PIGS methods to sample a single replica of a system (e.g. see Refs. [\onlinecite{Herdman2014},\onlinecite{Sarsa2000}]). Break-connect updates are those which break or reconnect a worldline at the central imaginary time-slice of a single replica and have been discussed in detail in Ref.~[\onlinecite{Herdman2014}].
 
Cross segment updates compose a new class that is required to ergodically sample the configuration space of the ratio method where worldlines of different replicas are connected at the center of the path in the ensemble. We introduce a \emph{cross-staging} update that chooses a bead in one replica with imaginary time slice index $p < P/2$ and another bead with $p\ge P/2$ in the other replica separated by $M < P$ time slices and attempts to perform a non-local staging update \cite{Sprik:1985bz} that can either connect or disconnect these worldlines across the break at the center of the path. Algorithmically:
\begin{enumerate}
    \item Choose a replica at random; we denote this as replica 1 and the other replica 2.  Choose a bead at time slice $p = P/2-1$ in replica 1 out of all worldlines that are either broken or cross-linked, and label this bead $b_-$; we denote the number of such beads as $n^A_1$. Follow the worldline back to $p{_<}=p-M/2$ and label this bead $b_<$.
    \item Define the number of beads in replica 2 at time slice $P/2$ that are in subregion $A$ as $n_2^A$. If $b_-$ is on a broken worldline, choose one of the $n_2^A$ beads and label it $b_{+}$. If $b_-$ is cross-linked between replicas, define $b_+$ to be the bead it is linked to. Follow the worldline of $b_+$ to time slice $p_{>} = p + M/2$ and label this bead as $b_>$.
\item Generate a new worldline segment between $b_<$ and $b_>$ of length $M$  with a weight given by the free particle propagator. Denote the updated beads about the center time-slice as $b_-'$ and $b_+'$. The updated segments in each replica are denoted by $(b_<,b_-')$ and $(b_+',b_>)$.
\item The acceptance probability $P_{\rm acc}$ of such an update depends on the ratio of the initial and final potential weights which we denote as $e^{-\delta U}$ as well as which of the four scenarios occur:
\begin{enumerate}
\item $b_+ \in A_b$ \& $b_+' \in A_b$:
\begin{equation}
    P_{\rm acc} = \frac{\rho_0\left(b_-,b_+\right)}{\rho_0\left(b_-',b_+'\right)}e^{-\delta U}
\end{equation}
\item $b_+ \in A_b$ \& $b_+' \in A_s$:
\begin{equation}
    P_{\rm acc} = n^A_2 \rho_0\left(b_-,b_+\right)e^{-\delta U}
\end{equation}
\item $b_+ \in A_s$ \& $b_+' \in A_b$:
\begin{equation}
    P_{\rm acc} = \frac{1}{n^A_2\rho_0\left(b_-',b_+'\right)}e^{-\delta U}
\end{equation}
\item $b_+ \in A_s$ \& $b_+' \in A_s$:
\begin{equation}
    P_{\rm acc} = e^{-\delta U}
\end{equation}
\end{enumerate}
\end{enumerate}
We note this update only operates on configurations with at least one broken or cross-linked worldline in each replica. We reject all updates which move $b_+$ into region $B$ as these configurations are ergodically sampled with break-connect updates ({\it e.g.} see 
Ref.~[\onlinecite{Herdman2014}]).

Another type of update we use for efficiency (although it is not generally required for ergodicity) is a cross-segment center-of-mass update. This update displaces the positions of all beads on a cross-linked worldline by a constant, and thus the acceptance rate only depends on the potential weights. This update is implemented identically to a conventional PIMC center-of-mass update \cite{Ceperley1995}, with the exception that if the bead at time slice $P/2+1$ is displaced out of $A_s$, the move is rejected.

 \subsection{Benchmarking}
% ---------------------------------------------------------------------------------

\begin{figure}[]
\begin{center}
\includegraphics[width=1.0\columnwidth]{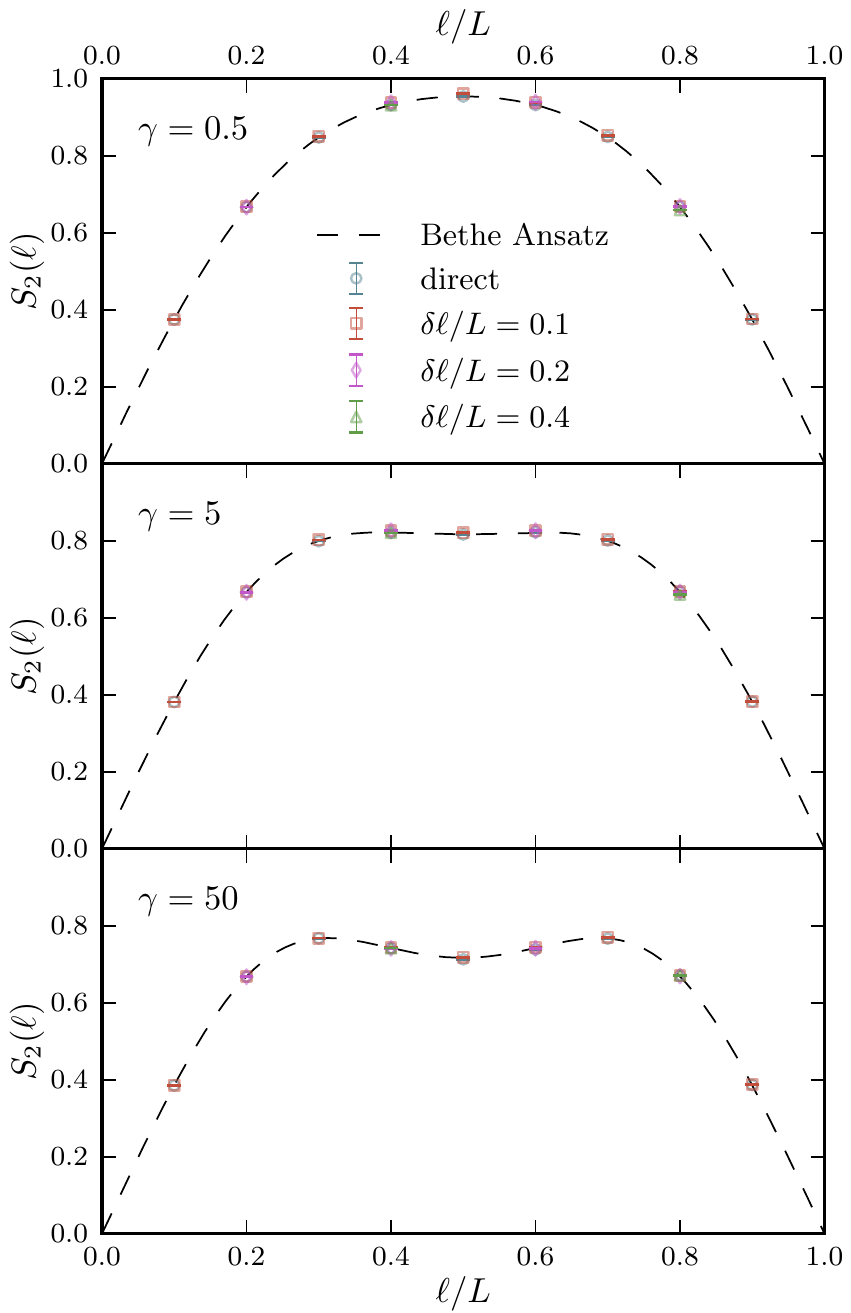}
\end{center}
\caption{(Color online) 2nd \ren entanglement entropy $S_2(\ell)$ of the ground state of $N=2$ Lieb-Liniger bosons as a function of spatial subsystem aspect ratio $\ell/L$ computed with QMC using both the direct and ratio method with step size $\delta\ell$. The interaction strength $\gamma$ is labeled on each plot, and increases from top to bottom. The solid lines are the exact results from the Bethe ansatz.}
\label{fig:N2Sva}
\end{figure}

 Having described the algorithmic details for continuous space worldlines, we now present proof of principle results for the ratio QMC method. To benchmark the QMC, we numerically compute $S_2(\ell)$ using the exact Bethe-Ansatz ground state wavefunction for a system of $N=2$ particles (see Appendix \ref{app:BetheAnsatz} for details). We take subsystem $A$ to be an interval of length $\ell$ and consider a variety of interaction strengths $\gamma$. For such a small systems size, numerical integration of the Bethe-ansatz ground state is tractable, so we can compare the QMC data to the exact ground state \ren entropies. We consider the ratio method using $K$ steps of size $\delta\ell$ such that $\Phi_k$, defined in \eqref{eq:Phik}, is computed at step $k$ with $\ell_k = k\delta \ell$ where $\delta \ell \equiv \ell/K$. The ratio method can then be employed to compute $S_2(\ell)$ from $K$ independent simulations. We compute $S_2(\ell)$ using the \emph{direct} (poorly scaling) QMC approach with a single interval, as well as using the ratio method for a variety of step sizes $\delta \ell$ for interaction strengths $\gamma=0.5,5,50$ and compare these to the Bethe-ansatz in \figref{fig:N2Sva}. In all cases, we find agreement with the exact ground state values.
%
% ------------------------------------------------------------------------------- 
% ------------------------------------------------------------------------------- 
%

%
% ------------------------------------------------------------------------------- 
\begin{figure}[t]
\begin{center}
\includegraphics[width=1.0\columnwidth]{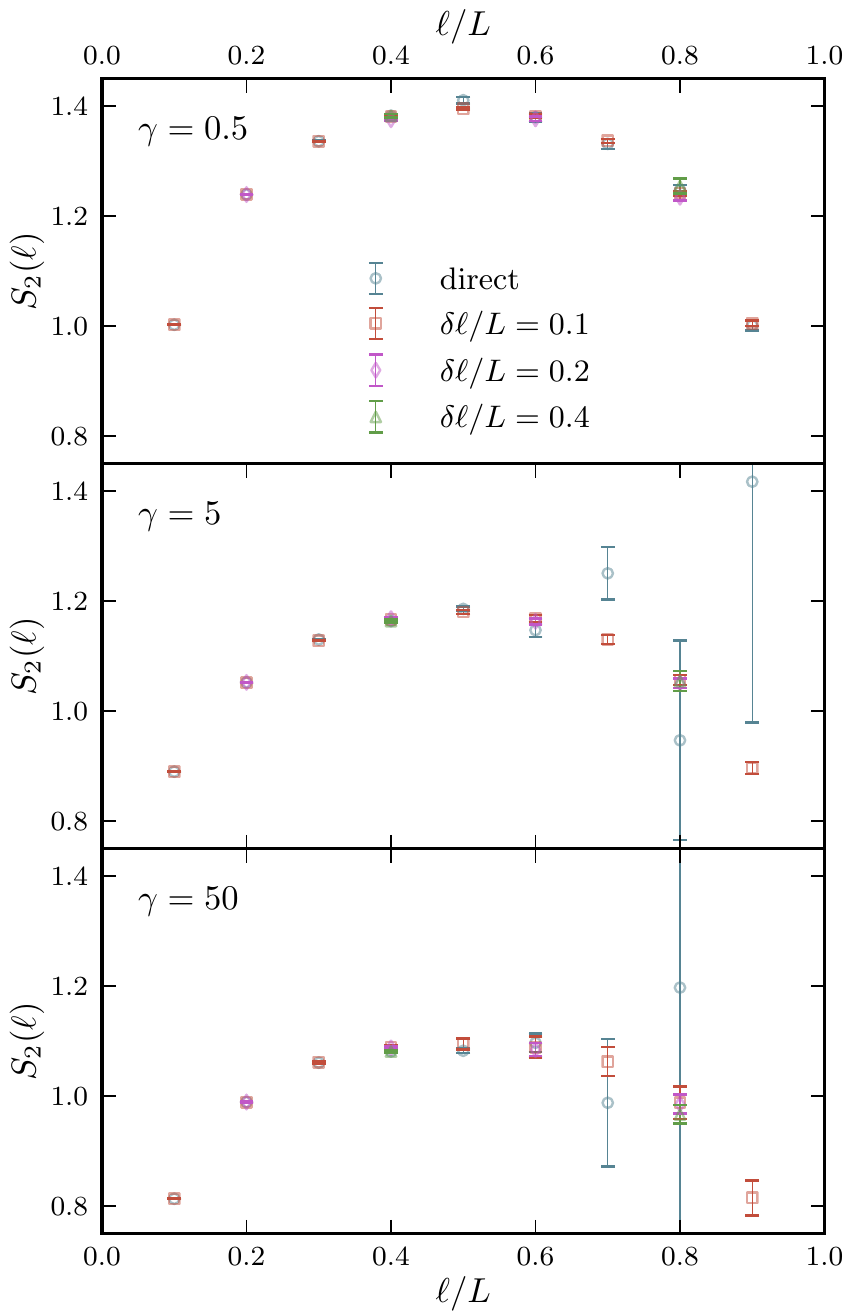}
\end{center}
\caption{(Color online) The 2nd \ren entanglement entropy $S_2(\ell)$ of the ground state of $N=8$ Lieb-Liniger bosons as a function of spatial subsystem aspect ratio $\ell/L$ computed with QMC using both the direct and ratio method with step size $\delta \ell$. The interaction strength $\gamma$ is labeled on each plot, and increases from top to bottom. The diverging statistical errors are indicative of the inefficiency of the direct method for larger subsystems.}
\label{fig:N8Sva}
\end{figure}
% ------------------------------------------------------------------------------- 
%

\figref{fig:N8Sva} shows the QMC results for a $N=8$ system with dimensionless interaction strengths $\gamma=0.5,5,50$ as a function of aspect ratio $\ell/L$ where it is not feasible to obtain the exact answer form the Bethe-ansatz. The diverging statistical uncertainties (for $\gamma=5,50$) for the direct QMC method for large $\ell$ demonstrates the inefficiency of the direct estimator. We find an improved statistical performance when employing the ratio method, as shown for several step sizes. The agreement between the ratio method for different steps sizes and the direct method (where it does not fail) provides confirmation of both its efficacy and accuracy.  In practice we find that the statistical performance of the direct SWAP estimator breaks down when the broken interval (of length $\delta \ell$) has order $4$ particles on average (although this is presumably strongly action and model dependent). Therefore in all subsequent results, we choose ratio method intervals that are sufficiently small to obtain this value. 

% ---------------------------------------------------------------------------------
 \section{\ren entanglement entropy in the Lieb-Liniger model}
% ---------------------------------------------------------------------------------

 Having suitably benchmarked the ratio method, we now present results of numerical calculations of the spatial \ren entanglement entropy of the ground state of the Lieb-Liniger model using the quantum Monte Carlo method described in Sec.~\ref{sec:QMC}. Using periodic boundary conditions, we consider system sizes up to $N=32$ at constant density, and bipartition the system into intervals of length $\ell$ and $L-\ell$. For each system size we consider the range $0.5 \leq \gamma \leq 50$ corresponding to moderate and strongly interactions regimes. For $N>8$ we use the ratio method, choosing a step size to be small enough for efficient performance of the estimator (as described above). In all cases we use a constant trial wavefunction at the ends of the imaginary time path and a sufficiently small finite time step $\tau$ and large imaginary time length $\beta$ such that systematic errors are smaller that the reported statistical errors.  See Appendix \ref{app:ConvQMC} for details of the $\tau$ and $\beta$ scaling of $S_2$.

To test the scaling predicted by CFT in \Eqref{eq:FSCFTc1} we fit the QMC data to the two parameter logarithmic scaling form
\begin{align}
S^{\rm{fit}}_2 \left(N,\ell\right) = \frac{c}{4} \log \Bigl[ 2 \pi n D\left(N,\ell\right) &\Bigr]+ c_2 \label{eq:FitForm} 
\end{align}
where $c$ and $c_2$ are two fit parameters. Additionally, we compare the QMC data for these interacting systems to the result for non-interacting bosons, where entanglement is generated purely from number fluctuations, which may be computed exactly from the result
\begin{align}
    S^{\rm{free}}_2 \left( N,\ell \right) &= \notag \\
    -\log &\left[ \sum_{n_A=0}^N  \binom{N}{n_A}^2  \left(\frac{\ell}{L}\right)^{2n_A} \left(1-\frac{\ell}{L}\right)^{2\left(N-n_A\right)} \right] \label{eq:freeEE}
\end{align}
(see e.g. Refs.~[\onlinecite{Herdman2014a}, \onlinecite{Simon2002}]).

\subsection{Moderate interaction regime $(\gamma < 1)$}
%
% ------------------------------------------------------------------------------- 
\begin{figure}[t]
  \centering
  \begin{tabular}{l}
    \subfigimg[width=\columnwidth]{(a)}{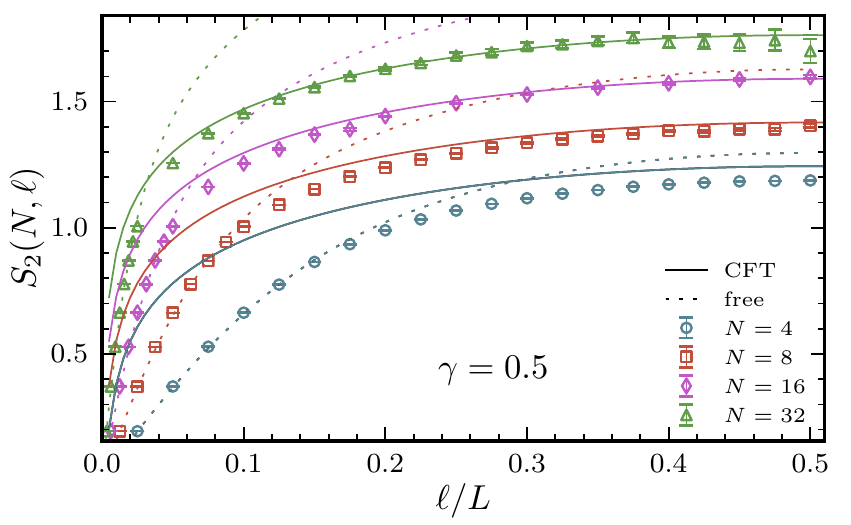} \\
    \subfigimg[width=\columnwidth]{(b)}{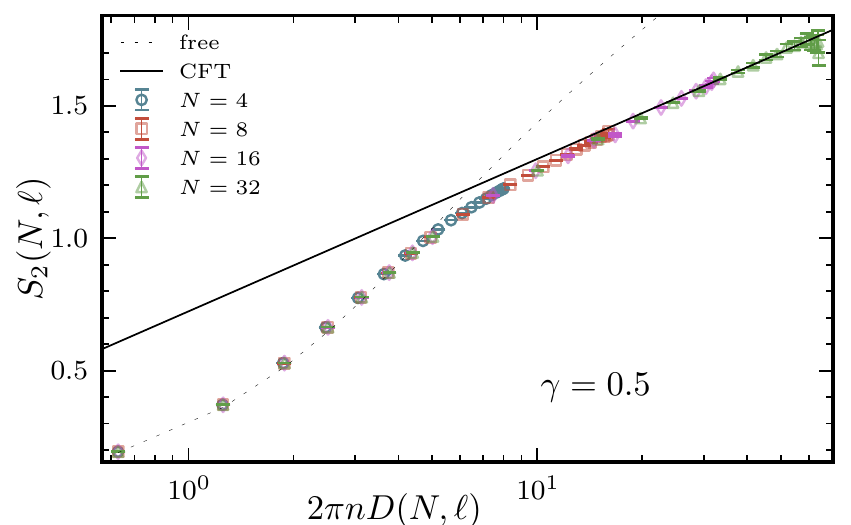}
  \end{tabular}
  \caption{(Color online) (a) The 2nd \ren entanglement entropy $S_2(\ell)$ of the ground state the Lieb-Liniger model with interaction strength $\gamma=0.5$ and systems sizes up to $N=32$ as a function of the subsystem aspect ratio $\ell/L$. (b) The same data is shown collapsed to a nearly pure function of chord length $D$. In both (a) \& (b) the solid lines represents a two parameter fit to the asymptotic scaling form given in \Eqref{eq:FitForm} while the dashed line represents the free boson result from \Eqref{eq:freeEE}. The dashed lines represents the free boson result from \Eqref{eq:freeEE}. The free boson results collapses nearly perfectly to a pure function of chord length within this regime, so the $N$ dependence is not visible in (b).}
\label{fig:SvlMod}
\end{figure}
% ------------------------------------------------------------------------------- 
%

\figref{fig:SvlMod} shows QMC data for $\gamma=1/2$, in the moderately interacting regime of the in the Lieb-Liniger model as functions of both aspect ratio $\ell/L$ and chord length $D(N,\ell)$. We find that the numerical data collapses onto a pure function of chord length. Fixing the leading coefficient to the CFT prediction $c = 1$, we perform a one-parameter fit for $c_2$ assuming the logarithmic scaling given in \Eqref{eq:FitForm}. As expected, due to finite size effects (e.g.~the CFT predicted power-law corrections), we only find a logarithmic fit on larger length scales.  This is clearly seen as the solid line in Fig.~\ref{fig:SvlMod}(b) represents a single fit to all the QMC data with $2\pi n D \gtrsim 20$. 

The dashed lines in \figref{fig:SvlMod} show the exact finite size free boson entanglement entropies given by \Eqref{eq:freeEE}. Within this moderately-interacting regime, the free boson EE collapses nearly perfectly to a pure function of chord length, and therefore there is no visible system size dependence in the dashed line of \figref{fig:SvlMod}(b). We find that for sufficiently small chord lengths, $(D\simeq \ell \lesssim \ell_0)$ the interacting results are well described by the free boson prediction, showing a clear deviation from the asymptotic logarithmic CFT scaling.   Thus, for $\ell_{\rm{int}}/\ell_0 = 2$ we find that for length scales $\ell \ll \ell_{\rm{int}}$ the EE is described by the free bosons result while for $\ell \gg \ell_{\rm{int}}$ it converges to the CFT scaling form. 

\begin{figure}[t]
\begin{center}
\includegraphics[width=1.0\columnwidth]{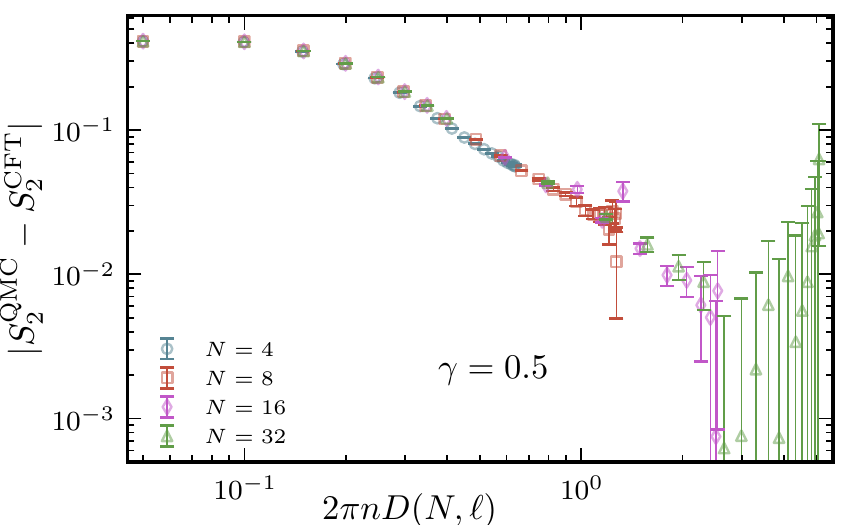}
\end{center}
\caption{(Color online) Deviations from the fit to the leading order logarithmic scaling of $S_2(\ell)$ for the Lieb-Liniger model with interaction strength $\gamma=0.5$ as a function of chord length. A reliable power-law fit to these corrections is not feasible with this data.}
\label{fig:CFT_cor}
\end{figure}

As mentioned in Sec.~\ref{sec:EECFT}, previous literature has demonstrated universal power-law corrections to \Eqref{eq:FitForm} in related models~\cite{Laflorencie2006,Calabrese2010c,Cardy2010,Xavier:2012en}. To investigate such possible corrections to the leading order scaling given in \Eqref{eq:FitForm}, in \figref{fig:CFT_cor} we have plotted the difference between the QMC data and a one parameter fit to \Eqref{eq:FitForm} with $c=1$ for $\gamma=0.5$. While this plot is suggestive of such power-law corrections, a reliable fit is not possible with this existing data. Our data for stronger interactions have even less visible corrections. Thus we leave a thorough analysis of these higher order corrections to the CFT scaling in the Lieb-Liniger ground state to future work.

%
% ------------------------------------------------------------------------------- 
% ------------------------------------------------------------------------------- 
%

\subsection{Strong interaction regime $(\gamma \gg 1)$}

\begin{figure}[t]
  \centering
  \begin{tabular}{l}
    \subfigimg[width=\columnwidth]{(a)}{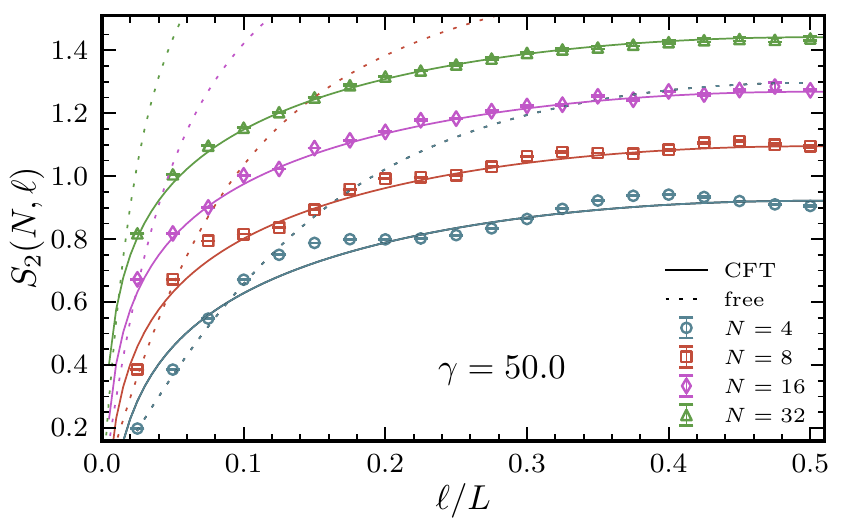} \\
    \subfigimg[width=\columnwidth]{(b)}{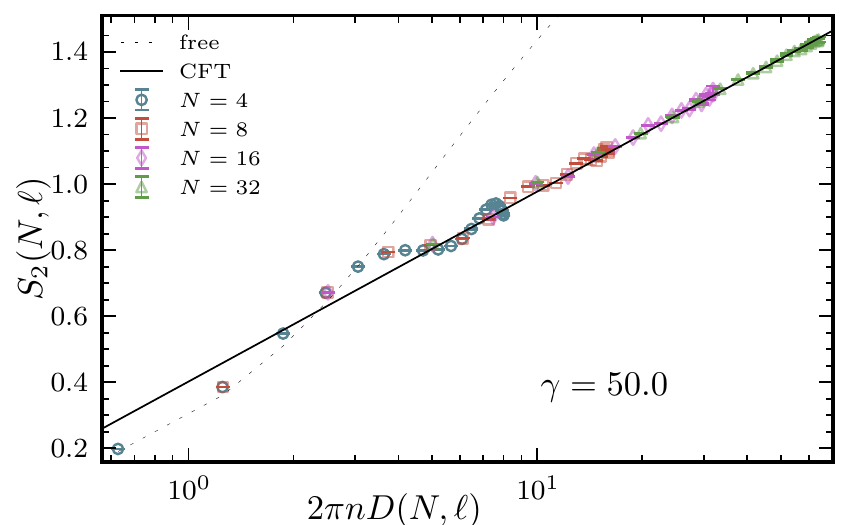}
  \end{tabular}
  \caption{(Color online) (a) The 2nd \ren entanglement entropy $S_2(\ell)$ of the ground sate of the Lieb-Liniger model with dimensionless interaction strength $\gamma=50$ and systems sizes up to $N=32$ as a function of subsystem aspect ratio $\ell/L$. (b) The same data is shown collapsed to a nearly pure function of chord length $D$. In both (a) and (b) the solid lines represent a two-parameter fit to the CFT prediction from \Eqref{eq:FitForm} while the dashed lines are free boson results from \Eqref{eq:freeEE}. The free boson results collapses nearly perfectly to a pure function of chord length within this regime, so the $N$ dependence is not visible in (b).} 
\label{fig:SvlStrong}
\end{figure}

\figref{fig:SvlStrong} shows $S_2$ for the ground state of the Lieb-Liniger model in the strongly interacting regime, with $\gamma=50$ as a function of both aspect ratio and chord length. Once again, we see convergence to the logarithmic CFT scaling (solid lines) at large length scales, and agreement with the free boson result (dashed lines) at short length scales. However in this case, the divergence from the free boson result occurs on shorter length scales than with weaker interactions -- this is consistent with the reduced interaction length scale $\ell_{\rm{int}}/\ell_0 = 0.02$ in this case. One strikingly different feature for strong interactions is the clear oscillations about logarithmic scaling that decay with the cord length $D$. Such oscillations have been previously observed in the $\alpha>1$ \ren EE of lattice spin models~\cite{Calabrese2010b}, dipolar lattice bosons \cite{Dalmonte:2011fm} and non-interacting fermions in the continuum~\cite{Calabrese2011a}, where they are related to power-law corrections to asymptotic logarithmic scaling.
%
% ------------------------------------------------------------------------------- 
% ------------------------------------------------------------------------------- 
%

\subsection{Scaling coefficients}

\begin{figure}[t]
  \centering
  \begin{tabular}{l}
    \subfigimg[width=\columnwidth]{(a)}{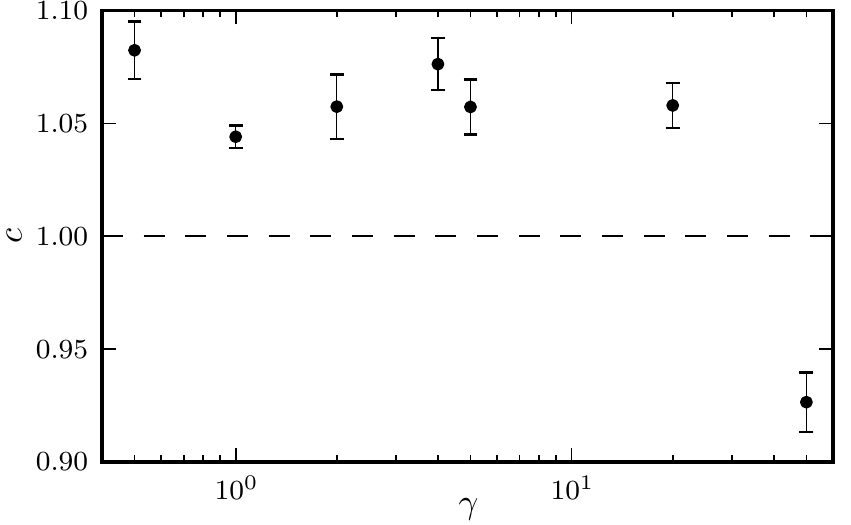} \\
    \subfigimg[width=\columnwidth]{(b)}{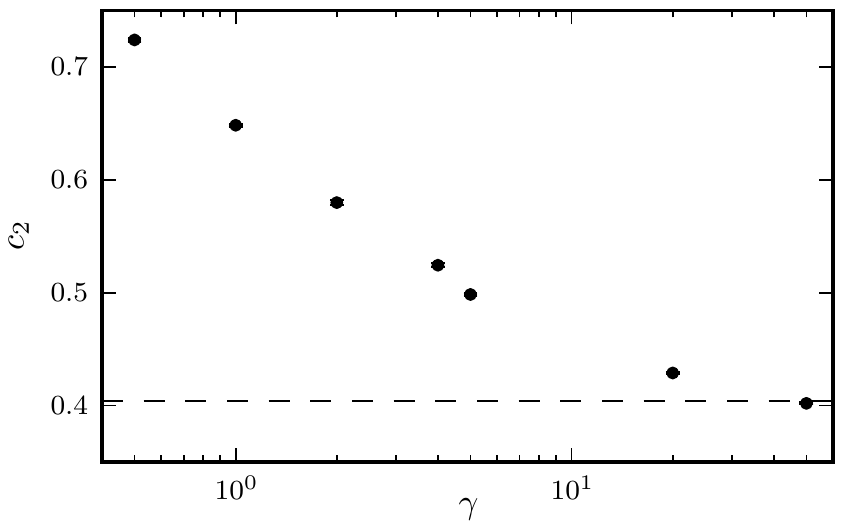}
  \end{tabular}
\caption{ The coefficients $c$ (a) and $c_2$ (b) of the leading order logarithmic scaling form of $S_2(\ell)$ given in \eqref{eq:FitForm} for the ground state of the Lieb-Liniger model determined by QMC as a function of the interaction strength $\gamma$. (a) The coefficient of the logarithmic scaling $c$ was determined by a two-parameter fit to the scaling form given in \eqref{eq:FitForm}. It is expected that $c=1$, the central charge of the associate conformal field theory.  The resulting deviations from the expected value on the order of $10\%$ are likely due to finite size effects. (b) The interaction dependence of the non-universal additive coefficient $c_2$ in the one-parameter fit (with $c=1$) to the logarithmic scaling given in \eqref{eq:FitForm} as a function of $\gamma$. The horizontal line corresponds to the free fermion value $c_2^{\rm FF} \simeq 0.404049$ which represents the Tonks-Girardeau limit of strongly interacting bosons. } 
\label{fig:cvg}
\end{figure}

For all interaction strengths $\gamma$, we can fit the asymptotic behavior of $S_2(\ell)$ to the logarithmic finite size scaling form of \Eqref{eq:FitForm} and extract the coefficients $c$ and $c_2$ using no prior knowledge of their values. 
\figref{fig:cvg}(a) shows the leading coefficient $c$ as function $\gamma$ extracted  in this way.
%
% ------------------------------------------------------------------------------- 
% ------------------------------------------------------------------------------- 
%
From Luttinger liquid theory, we expect $c=1$, as $c$ is the central charge of the CFT and our finite size QMC data agrees with this prediction to within $10\%$.  The residual discrepancy of the numerically extracted value of the central charge $c$ is likely due to finite size effects; indeed this analysis ignores the possible power-law corrections inferred in \Eqref{eq:FSCFT}. Additional complications arise in the fitting procedure due to the oscillatory nature of $S_2$ for large interactions $\gamma$. 

In an attempt to mitigate these residual finite size effects while extracting an estimate of the sub-leading constant $c_2$, we now fix $c=1$ and perform a one-parameter fit, which is shown in \figref{fig:cvg}(b). In the limit $\gamma\rightarrow\infty$, we expect $c_2$ to converge to the free-fermion value which is shown as a horizontal line in \figref{fig:cvg}(b); indeed we find $c_2(\gamma)\approx c_2^{\rm{FF}}\simeq 0.404049$.

% ---------------------------------------------------------------------------------
 \section{Discussion} 
% ---------------------------------------------------------------------------------
 In this paper, we have numerically studied the finite size scaling of the 2nd \ren entanglement entropy of the ground state of the Lieb-Liniger model of contact interacting bosons in the one dimensional spatial continuum. We find that the asymptotic scaling of $S_2$ agrees with the predicted logarithmic scaling of conformal field theory, with a leading coefficient consistent with central charge $c=1$. We note that the uncertainty of $\sim10\%$ is not much larger than that inferred from a recent continuous space matrix product state study on the same model \cite{Stojevic:2015dj} which employed a sort of effective finite size scaling based on the bond dimension.  Systematic and statistical errors could be further reduced by pushing our simulations to larger values of $N$.

We have measured the non-universal sub-leading constant as a function of the dimensionless interaction strength $\gamma$ and find it to be a monotonically decreasing function of $\gamma$ that converges to the free-fermion value for sufficiently strong interactions.  This behavior is consistent with the $c_2$ dependence found for increasing anisotropy in the XXZ model \cite{Calabrese2010b} which corresponds to stronger nearest neighbor repulsion in the equivalent itinerant hardcore boson model.  On shorter length scales there is a crossover to free boson behavior, where the crossover length scale depends on interaction strength.

This study demonstrates how algorithmic advances are important for the study of universal scaling of entanglement entropy in interacting itinerant bosons.  In the case of the Lieb-Liniger model with moderate interaction strength, reliable finite-size scaling data plays a crucial role in identifying possible power-law corrections to the leading order logarithmic scaling expected for the low-energy effective conformal field theory description.  We expect that high-precision quantum Monte Carlo simulations such as this will be vital in the continuing exploration of entanglement entropy in systems of itinerant particles in the future.

% ---------------------------------------------------------------------------------
 \section{Acknowledgments}
% ---------------------------------------------------------------------------------
We are indebted to Erik Tonni and John Cardy for their insights related to scaling corrections in conformal field theory.  This research was supported in part by the National Science Foundation under Grant No. NSF PHY11-25915. Additionally, we thank the Natural Sciences and Engineering Research Council of Canada for financial support. Computations were performed on the Vermont Advanced Computing Core supported by NASA (NNX-08AO96G) as well as on resources provided by the Shared Hierarchical Academic Research Computing 
Network (SHARCNET).

% ---------------------------------------------------------------------------------
\appendix
% ---------------------------------------------------------------------------------

% ---------------------------------------------------------------------------------
\section{\ren entropy from the Bethe Ansatz}
\label{app:BetheAnsatz}
% ---------------------------------------------------------------------------------

Here we briefly describe the method used to compute the \ren entropy of the ground state of the Lieb-Liniger model using the Bethe Ansatz. For $N=2$, the Bethe ansatz wave function is:
\begin{equation}
\Psi \left(x_1,x_2\right) = A_{1,2} e^{ i \left( k_1 x_1 + k_2 x_2 \right)}+ A_{2,1} e^{ i \left( k_2 x_1 + k_1 x_2 \right)} \notag
\end{equation}
for $x_1\leq x_2$, where $k_1$ \& $k_2$ are real \emph{quasi-momenta} with $k_1 < k_2$, and the $A$'s are complex coefficients. We restrict ourselves to the $k_1=-k_2 = -k$ state and choose the coefficients such that $\Psi$ is an energy eigenstate of \Eqref{eq:HLieb} with energy $2\lambda k^2$; this fixes $\Psi$ to be 
\begin{align}
\Psi \left(x_1,x_2\right) = \frac{1}{\sqrt{Z}} &\biggl( 2 \cos \bigl[k\left(x_2-x_1 \right) \bigr] \notag\\
&+ \frac{1}{k\li} \sin \bigl[k\left(x_2-x_1 \right) \bigr] \biggr)  \label{eq:psiBA}
\end{align}
where $Z$ is a normalization factor:
\begin{equation}
Z = \frac{1}{4} \left( \frac{L}{k \li} \right)^2 \left( 1 + 4 \frac{\li}{L} + 4 \left( k \li \right)^2 \right) \notag
\end{equation}
and $k$ is a solution to the Bethe equation:
\begin{equation}
k L = \pi n -2\arctan\left[ 2 k \li \right]. \label{eq:Bethe}
\end{equation}
\Eqref{eq:Bethe} can be solved numerically for $k$ and using this value of $k$ in \Eqref{eq:psiBA} provides the exact ground state wave function. 

We can write the reduced density matrix of an interval of length $\ell$ as
\begin{equation}
\rho_A = P_0 \rho_0 + P_1 \rho_1 + P_2 \rho_2 \notag
\end{equation}
where $P_n$ is the probability of finding $n$ particles in $A$ and $\rho_n$ is the reduced density matrix projected onto the $n$ particle subspace of $\rho_A$. The purity of $\rho_A$ may then be written as:

\begin{align}
\Tr \rho_A^2 &= P_0^2\Tr \rho_0 ^2 + P_1^2\Tr \rho_1 ^2 +P_2^2 \Tr \rho_2 ^2 \notag\\
& = P_0^2 + P_1^2\Tr \rho_1 ^2 + P_2^2 \label{eq:TrrhoA2}
\end{align}
$P_n$ and $\rho_1$ may be computed analytically and are found to be the following:

\begin{widetext}
\begin{align}
P_0 =& \int_{\ell} ^{L}dx_1 \int_{x_1} ^{L}dx_2 \Psi^*\left(x_1,x_2\right) \Psi \left(x_1,x_2\right) \notag \\
=& \frac{1}{4 k^2 Z} \biggl\{ \left(4-\frac{1}{k^2 \li^2}\right) \sin ^2\left[ k \left(L-\ell\right)\right] + \left(\frac{L-\ell}{\li}\right)^2 \Bigl(4 k^2 \li ^2 +1+4 \frac{\li}{L-\ell} \Bigr) -\frac{2} {k\li}  \sin \left[2 k\left(L-\ell\right)\right] \biggr\} \label{eq:P0}\\
P_2 =& \int_{0} ^{\ell}dx_1 \int_{x_1} ^{\ell}dx_2 \Psi^*\left(x_1,x_2\right) \Psi \left(x_1,x_2\right) \notag \\
=& \frac{1}{4 k^2 Z} \biggl\{ \left(4 -\frac{1}{k^2 \li ^2}\right) \sin ^2\left[\ell k\right] + \left(\frac{\ell}{\li}\right)^2 \left(4 k^2 \li ^2+1+4 \frac{\li}{\ell} \right)-\frac{2}{k \li }  \sin \left[2 \ell k\right] \biggr\} \label{eq:P2}
\end{align}
\begin{align}
P_1 \rho_1 \left(x,x'\right)  =& \int_{\ell} ^{L}dx'' \Psi^*\left(x,x''\right) \Psi \left(x',x''\right) \notag \\
=&\frac{1}{4 k Z}\Bigl\{ -8 k \ell \cos \left[k (x-x')\right]-4 \sin \left[ k (2 \ell-x-x')\right] +\frac{4}{k \li} \cos\left[k (2 \ell-x-x')\right]\notag\\
&\quad\qquad+\frac{1}{k^2\li^2}\sin\left[k (2 \ell-x-x')\right]- \frac{2\ell}{k\li^2} \cos\left[k (x-x')\right]+8 k L \cos\left[k (x-x')\right]+4  \sin\left[k (2 L-x-x')\right]\notag\\
&\quad\qquad-\frac{4}{k \li}  \cos\left[k (2 L-x-x')\right]-\frac{1}{k^2\li^2}\sin\left[k (2 L-x-x')\right]+\frac{2L}{k\li^2} \cos \left[k (x-x')\right] \Bigr\} \label{eq:rho1}
\end{align}
\end{widetext}
Using \Eqref{eq:rho1}, $P_1^2 \Tr \rho_1^2$ may be computed by numerical integration:
\begin{equation}
P_1^2 \Tr \rho_1^2 = \int_{0} ^{\ell}dx \int_0^{\ell}dx' P_1^2\rho_1(x,x')^2. \label{eq:Trrho12}
\end{equation}
Finally, $S_2(\ell)$ is computed from \Eqref{eq:TrrhoA2}, using \Eqref{eq:P0}, \Eqref{eq:P2} \& \Eqref{eq:Trrho12}.
% ---------------------------------------------------------------------------------
\section{Convergence with quantum Monte Carlo parameters}
\label{app:ConvQMC}
% ---------------------------------------------------------------------------------

\begin{figure}[t!]
\begin{center}
\includegraphics[width=1.0\columnwidth]{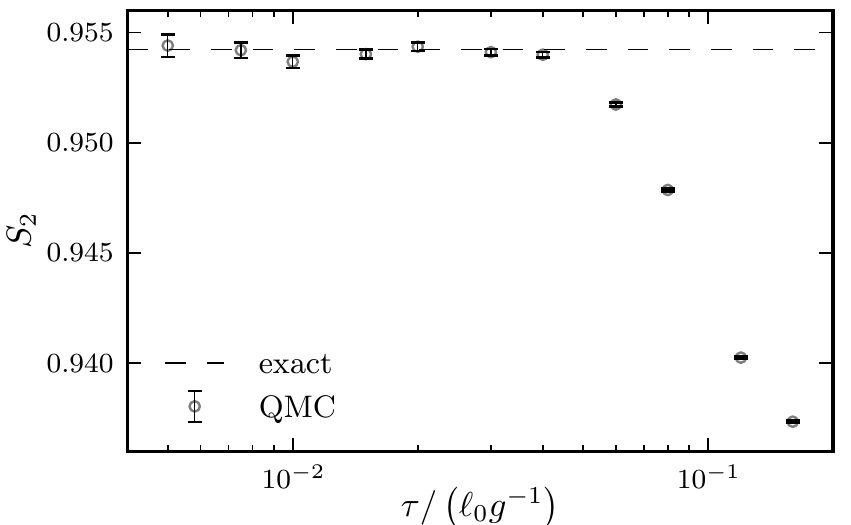}
\end{center}
\caption{Quantum Monte Carlo convergence of the 2nd \ren entanglement entropy $S_2(a)$ with decreasing imaginary time step-size $\tau$ for a symmetric partition with $\ell/L = 1/2$
for $N=2$ Lieb-Liniger bosons with $\gamma=1/2$ and $\beta/(Lg^{-1}) = 0.64$. The dashed line corresponds to the exact Bethe-ansatz value.}
\label{fig:N2Svt}
\end{figure}

In this appendix we demonstrate the convergence of the \ren entropies calculated with quantum Monte Carlo with the imaginary time length $\beta$ and finite-time step $\tau$. Discrete imaginary-time worldline QMC methods introduce a controlled systematic error due to the finite imaginary time step size $\tau$. \figref{fig:N2Svt} shows the convergence of the QMC data to the exact Bethe-ansatz value for an $N=2$ system of Lieb-Liniger bosons.
%
% ------------------------------------------------------------------------------- 
% ------------------------------------------------------------------------------- 
%

\begin{figure}[t!]
\begin{center}
\includegraphics[width=1.0\columnwidth]{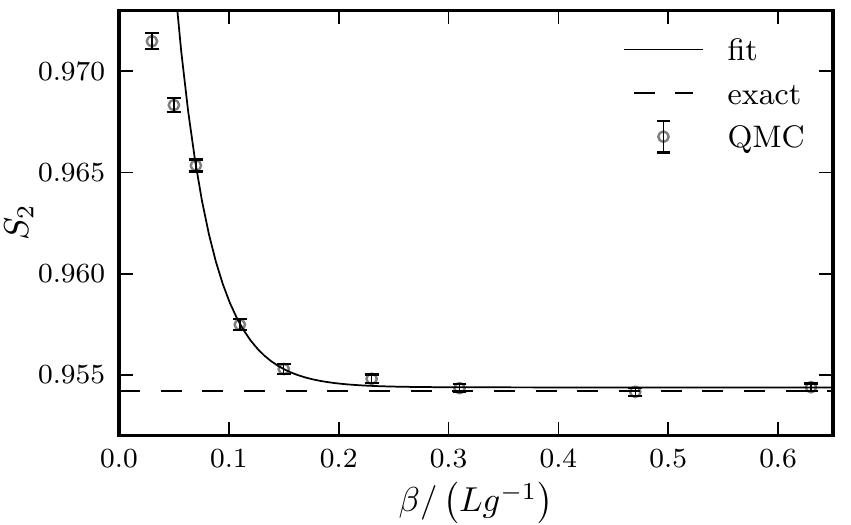}
\end{center}
\caption{Quantum Monte Carlo convergence of the 2nd \ren entanglement entropy $S_2(a)$ with decreasing imaginary time length $\beta$ for a symmetric partition with $\ell/L = 1/2$
for $N=2$ Lieb-Liniger bosons with $\gamma=1/2$ and $\tau/(\ell_0g^{-1}) = 0.02$.  The solid line is a fit to \Eqref{eq:S2Scalebeta} with $S_0 = 0.9544(1)$, $c_\beta = 0.10(2)$ and $\delta/(gL^{-1}) = 33(2)$. The dashed line corresponds to the exact Bethe-ansatz value.}
\label{fig:N2SvP}
\end{figure}

Path integral ground state based QMC methods also introduce a systematic error based on a finite imaginary-time length $\beta$. We characterize this error by scaling $\beta$ and fitting to the exponential:
\begin{equation}
S \left( \beta \right) = S_0 + c_\beta~e^{-\delta \beta}
\label{eq:S2Scalebeta}
\end{equation}
with prefactor $c_\beta$ and $\delta$ has units of energy. \figref{fig:N2SvP} shows the convergence of the QMC data to the exact Bethe-ansatz value for an $N=2$ system.
%
% ------------------------------------------------------------------------------- 
% ------------------------------------------------------------------------------- 
%
\FloatBarrier
\bibliography{LiebLinigerSpatEE}

\end{document}